%% file: paper.tex
\documentclass[conference]{IEEEtran}
\IEEEoverridecommandlockouts
\usepackage{cite}
\usepackage{amsmath,amssymb,amsfonts}
\usepackage{graphicx}
\usepackage{textcomp}
\usepackage{xcolor}
\usepackage{microtype}
\usepackage{xspace}
\usepackage{subcaption}
\usepackage{wrapfig}
\usepackage{algorithm}
\usepackage{amsthm}
\usepackage{verbatim}
\usepackage{mathtools}
\usepackage{soul}
\usepackage[shortlabels]{enumitem}
\usepackage{bbm}
\usepackage{multirow}
\usepackage{algpseudocode}
\usepackage{amsmath}
\usepackage{hyperref}
\usepackage{booktabs}
\usepackage{todonotes}

\usepackage[normalem]{ulem}
\usepackage{xcolor}

\newcommand{\newtext}[1]{{\color{black}#1}}

\usepackage{tikz}
\usepackage{textcomp}
\usepackage{hyperref}
\usepackage{lipsum}

\newcommand\copyrighttext{%
  \footnotesize \textcopyright 2025 IEEE. Personal use of this material is permitted.
  Permission from IEEE must be obtained for all other uses, in any current or future
  media, including reprinting/republishing this material for advertising or promotional
  purposes, creating new collective works, for resale or redistribution to servers or
  lists, or reuse of any copyrighted component of this work in other works. DOI: \href{https://doi.org/10.1109/ISM66958.2025.00035}{10.1109/ISM66958.2025.00035}
  }
\newcommand\copyrightnotice{%
\begin{tikzpicture}[remember picture,overlay]
\node[anchor=south,yshift=10pt] at (current page.south) {\fbox{\parbox{\dimexpr\textwidth-\fboxsep-\fboxrule\relax}{\copyrighttext}}};
\end{tikzpicture}%
}

\let\oldsout\sout
\renewcommand{\sout}[1]{{\color{red}\oldsout{#1}}}

\def\BibTeX{{\rm B\kern-.05em{\sc i\kern-.025em b}\kern-.08em
    T\kern-.1667em\lower.7ex\hbox{E}\kern-.125emX}}
\begin{document}

\title{Secure AI-Driven Super-Resolution for Real-Time Mixed Reality Applications}




\author{
\IEEEauthorblockN{Mohammad Waquas Usmani\IEEEauthorrefmark{1},
Sankalpa Timilsina\IEEEauthorrefmark{2},
Michael Zink\IEEEauthorrefmark{1}, and
Susmit Shannigrahi\IEEEauthorrefmark{2}}

\IEEEauthorblockA{\IEEEauthorrefmark{1}\textit{University of Massachusetts Amherst}, Massachusetts, USA \\
Email: mohammadwaqu@umass.edu, zink@ecs.umass.edu}

\IEEEauthorblockA{\IEEEauthorrefmark{2}\textit{Tennessee Technological University}, Tennessee, USA \\
Email: stimilsin43@tntech.edu, sshannigrahi@tntech.edu}
}

\maketitle
\copyrightnotice

\begin{abstract}
\input{0-abstract}
\end{abstract}

\begin{IEEEkeywords}
Point Clouds, Mixed Reality, \newtext{Machine Learning based Super-Resolution}, Attribute-Based Encryption, Network Performance
\end{IEEEkeywords}

\input{1-Introduction}
\input{2-Background}
\input{3-Architecture}
\input{4-Evaluation}
\input{5-Discussion}
\input{6-Conclusion}

\section*{Acknowledgment}

This material is based upon work supported by the National Science Foundation under grants \#2319962 and \#2126148.

\bibliographystyle{IEEEtran}
\bibliography{refs}

\end{document}

%% file: 0-abstract.tex
Immersive formats such as 360° and 6DoF point cloud videos require high bandwidth and low latency, posing challenges for real-time AR/VR streaming. This work focuses on reducing bandwidth consumption and encryption/decryption delay, two key contributors to overall latency. We design a system that downsamples point cloud content at the origin server and applies partial encryption. At the client, the content is decrypted and upscaled using an ML-based super-resolution model. \newtext{Our evaluation demonstrates a nearly linear reduction in bandwidth/latency, and encryption/decryption overhead with lower downsampling resolutions, while the super-resolution model effectively reconstructs the original full-resolution point clouds with minimal error and modest inference time.}

%% file: 1-Introduction.tex
\section{Introduction}
New immersive multimedia technologies and formats are gaining popularity. Such formats include 360° videos (3DoF), six-degrees-of-freedom (6DoF) point clouds, and 3D objects for augmented reality applications. This trend is driven by the availability of devices such as Apple Vision Pro and Meta Quest 3, allowing users to consume this content in high quality.

A significant challenge for 3DoF and 6DoF streaming that does not exist for regular video streaming is motion-to-photon latency, the delay between viewer movement and the updated visuals displayed on a device. High latency can cause motion sickness and reduce immersion in VR and AR\cite{shannigrahi2020next}. Low latency ensures a smoother, more responsive experience.
Streaming applications are also bandwidth-intensive 
since high bandwidth usage may lead to congestion and packet loss, which increase the overall latency.


In our earlier work, we showed that using in-network caching and data reuse can cut down network latency~\cite{10.1145/3712676.3714450,usmani2025securingimmersive360video}. In this paper, we reduce latency by introducing a novel method that combines super-resolution and Attribute-\newtext{B}ased Encryption (ABE)~\cite{CPABE}. This approach provides AR/VR streaming applications with the opportunity to trade off between upscaling via \newtext{super-resolution}~(SR)~\cite{yu2018pu} at the client, data encryption and decryption, and the amount of data transferred over the network. 



In this work, we build a system that processes point cloud video streams by downsampling them at the origin server to reduce transmission overhead. To evaluate the associated trade-offs, the system provides multiple downscaled versions of the content at the server. 
This content is then partially encrypted before transmission to the client. 
At the client, data is decrypted and upscaled via ML-based super-resolution. 

We evaluate our approach using a point-cloud data set that consists of scenes from offices and living rooms. We use a subset of these scenes to train an SR model that is used at the client for upsampling. In addition, we make use of the ABE-based partial encryption approach we developed for point-cloud streaming~\cite{usmani2025securingimmersive360video}. The performance of this system is evaluated based on standard QoE metrics for point-clouds.

\newtext{Our contribution is that we introduce a novel perspective by integrating \textit{security} into the ML-based super-resolution pipeline for point clouds. Specifically, we combine ABE with point cloud downsampling and ML-based SR. This framework reduces bandwidth usage through downsampling while simultaneously lowering encryption-decryption overhead, an aspect overlooked in prior research. We further develop and train an efficient AI/ML model capable of reconstructing full-resolution point clouds from downsampled data with high fidelity. Experimental results show a nearly linear reduction in network latency, bandwidth consumption, and encryption-decryption time with increased downsampling, while the proposed super-resolution model restores the original resolution with minimal reconstruction error and modest inference time.}






%% file: 2-Background.tex
\section{Background}

\begin{figure*}[!ht]\centering
  \includegraphics[width=0.75\textwidth]{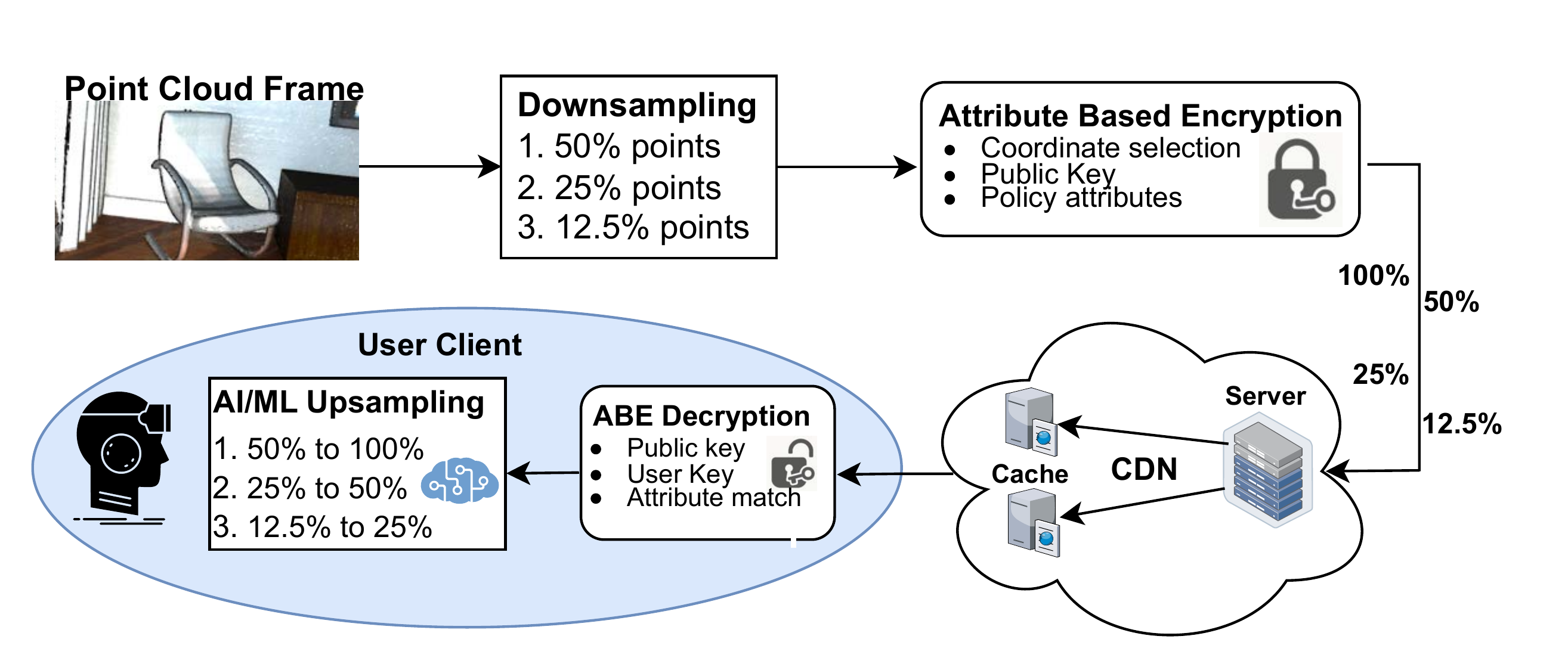} 
  \caption{System architecture. Point clouds are downsampled from full-100\% resolution to 50\% and 25\%, and 12.5\%, encrypted using ABE, stored at the origin, and delivered via CDN. Clients decrypt and upsample the frames using an AI/ML model.} 
  \label{sys-arch} 
\end{figure*}

\subsection{3D Point Clouds}
\label{subsec:pointcloud-def}
Point clouds represent a collection of discrete points in three-dimensional space, where each point is characterized by its spatial coordinates \((x, y, z)\) and may include additional properties such as color \((r, g, b)\) or surface normals \((n_x, n_y, n_z)\). Combined, these points reconstruct the geometry of an object or scene and are commonly stored in the Polygon File Format (\texttt{.ply})~\cite{ply}. Unlike conventional 2D image or video data, point clouds provide a flexible, spatially explicit representation of the physical environment. When captured sequentially over time, these frames form a volumetric video that serves as a core building block for mixed reality (MR) systems that integrate elements of both virtual and augmented environments. Such representations enable 6DoF for immersive user interaction and spatial navigation~\cite{10.1145/3394171.3413639}, offering a level of realism and interactivity that extends far beyond the three degrees of freedom typical of traditional 360° video experiences~\cite{usmani2025securingimmersive360video}.

\subsection{Related Work on Point Cloud Super-Resolution}
Super-resolution (SR) using deep learning was first explored for images~\cite{Dong2014LearningAD,8099502} and later extended to conventional 2D videos ~\cite{7410425,caballero2017realtimevideosuperresolutionspatiotemporal}. The demand for enhancing resolutions eventually led to its application in emerging 3D formats such as point clouds. The pioneering work in learning-based point cloud super-resolution is \textit{PU-Net}\cite{yu2018pu}, which introduced one of the first neural network architectures for point cloud upsampling. Building on this foundation,\cite{10503354} extended SR techniques to LiDAR and large-scale 3D scenes. Other recent studies~\cite{10336534} have expanded point cloud SR beyond geometric spatial coordinates to also enhance per-point color attributes, while~\cite{10.1007/978-3-030-58529-7_44} introduced a geometrically aware upsampling framework.

In addition to learning-based approaches, several optimization-based methods have been proposed. For example,~\cite{9682535,8803560} formulate point cloud upsampling and surface reconstruction as optimization problems, without relying on data-driven learning.

There have also been streaming-oriented studies on applying learning based SR to point clouds for volumetric video streaming~\cite{10.1145/3446382.3448663,278306,wang2025volutefficientvolumetricstreaming}. The \textit{YuZu} framework~\cite{278306} represents one of the most comprehensive efforts to integrate SR into volumetric video streaming. Their approach transmits low-resolution point cloud segments and reconstructs higher-resolution versions at the client side using SR models. The system incorporates adaptive bitrate (ABR) control within a DASH-based pipeline and performs evaluations using a client-server setup. Additionally, they introduce a Quality of Experience (QoE) model to assess perceptual quality and demonstrate that SR-based reconstruction achieves significantly better QoE compared to viewport-adaptive methods. More recently, an extension of this work,  \textit{VoLUT}~\cite{wang2025volutefficientvolumetricstreaming}, introducing a lookup-table (LUT)-based SR approach to further 
improve efficiency during upscaling.

While prior works have extensively investigated SR-based bandwidth optimization for volumetric and point cloud streaming, our approach introduces a novel perspective by integrating \textit{security} into this pipeline. Specifically, we combine 
ABE with point cloud downsampling and ML-based SR. Our framework reduces bandwidth usage through downsampling and lowers encryption-decryption overhead, addressing a dimension that has received little attention in prior work. Moreover, we train an AI/ML model capable of efficiently reconstructing full-resolution point clouds from downsampled data. Although our current focus is on evaluating downsampling, ABE, and reconstruction performance, future work will extend this framework to a full streaming setup with ABR, DASH integration, and QoE evaluation similar to~\cite{278306}.

\subsection{Related Work on Attribute-Based Encryption}
\label{sec:abe-intro}
Secure and flexible access control is a key requirement for collaborative or distributed applications, such as shared 3D or MR data platforms, where users from different organizations or regions access common content. Attribute-Based Encryption (ABE)~\cite{CPABE} addresses this need by enabling fine-grained, policy-driven access control directly at the cryptographic layer.


\paragraph{Access Control Scenario}

\newtext{Consider an MR data-sharing platform that limits access to certain 3D datasets based on user attributes. The data owner defines a policy allowing only users with \emph{Role: Researcher} and either \emph{Affiliation: University~X} or \emph{Region: Europe}. In ABE, this rule is embedded in the ciphertext during encryption with public key~$PK$. Each user holds a private key~$SK$ encoding their attributes. For example, Alice has ${\texttt{Role: Researcher}, \texttt{Affiliation: UnivX}}$, while Bob has ${\texttt{Role: Student}, \texttt{Region: Europe}}$. On decryption, the system checks if the user’s attributes satisfy the ciphertext policy. Since Alice’s attributes meet the access condition, she can decrypt the dataset, whereas Bob cannot. This approach enforces fine-grained, cryptographic access control without per-user key management.}

\paragraph{Key Revocation and Updates}
ABE further supports dynamic key and attribute management. If a user’s status changes (for example, Alice leaves the research group), her private key~$SK$ can be revoked or allowed to expire without re-encrypting the data. Time-bound revocation can also be enforced by embedding expiration attributes or validity periods in keys, as demonstrated in~\cite{reddick2022wip, reddick2021design, reddick2022case}. Administrators can periodically refresh or revoke these attributes to maintain continuous security.






Prior work~\cite{10.1007/978-3-319-32689-4_26,10.1145/3712676.3714450,usmani2025securingimmersive360video} has explored the use of ABE in video streaming, primarily focusing on conventional 2D and 360$^\circ$ video content. The latter studies~\cite{10.1145/3712676.3714450,usmani2025securingimmersive360video} demonstrated through real-time streaming evaluations that ABE over HTTP can reduce cache CPU load while maintaining Quality of Experience (QoE) comparable to HTTPS. In particular,~\cite{usmani2025securingimmersive360video} extended this concept by integrating selective frame-level encryption with viewport-adaptive protection.

More specifically for point cloud data, \cite{10.1145/3704413.3765298} introduced a selective coordinate encryption mechanism using ABE, enabling encryption of different subsets of spatial coordinates \(x, y, z\) to balance computational cost against the desired level of obfuscation. In this work, we build upon and adapt that mechanism to secure point cloud content, as further discussed in Sect.~\ref{ABE-architeture}.

%% file: 3-Architecture.tex
\section{Secure AI-Driven SR Distribution Architecture} \label{sec:arhcitecture}

\newtext{Fig. \ref{sys-arch} shows the overall system architecture in a real-world deployment scenario. In this scenario,} each point cloud frame in the video sequence undergoes a multi-stage processing pipeline. First, every frame is downsampled by iteratively removing every second point, producing reduced-resolution versions at 50\%, 25\%, and 12.5\% of the original point count. Next, the downsampled frames are secured using ABE at the chosen coordinate granularity. The encrypted frames are stored on the origin server and distributed through a Content Delivery Network (CDN).
On the client side, users retrieve the encrypted frames, perform decryption, and then apply an AI/ML-based upsampling model to reconstruct higher-resolution point clouds for playback. This section details the processing pipeline, including downsampling, ABE, and AI/ML-based SR. \newtext{Note that we do not perform any transport optimization or caching; our focus in this work is solely on encryption/decryption and up- and downsampling.}


\subsection{Downsampling of Point Cloud Frames}
To simplify the evaluation, each point cloud is downsampled by removing every second point, effectively halving the total number of points and reducing the data size proportionally. This process is repeated iteratively to generate three additional resolutions of each frame with progressively smaller sizes: $100\% \;\rightarrow\; 50\% \;\rightarrow\; 25\% \;\rightarrow\; 12.5\%.$ The downsampled versions are then used to analyze data size/bandwidth reduction and the effectiveness of subsequent upsampling by the deployed AI/ML model~(see Sect. ~\ref{evaluation}).

\subsection{ABE for Securing Point Clouds} \label{ABE-architeture}
\newtext{To secure the point cloud video streams, we adopt the attribute-based selective coordinate encryption method from our prior work~\cite{10.1145/3704413.3765298}. This approach supports encryption at different granularities, encrypting all $X$, $Y$, and $Z$ coordinates, only $X$ and $Y$, or just $X$. The selected coordinates are extracted and encrypted using ABE, with the resulting ciphertext embedded in the point cloud frame as metadata while the original coordinates are removed. During decryption, the metadata is parsed, the encrypted values are recovered, and the original frame is reconstructed by reinserting the decrypted coordinates into their original positions.}


\newtext{In this work, we encrypt all $X$, $Y$, and $Z$ coordinates to evaluate the computational savings achieved through downsampling, while prior work~\cite{10.1145/3704413.3765298} analyzed the trade-offs of selective encryption in detail. The use of ABE ensures data confidentiality both in transit and at rest, while avoiding per-client encryption overhead~\cite{10.1145/3712676.3714450}. This design also preserves the caching and distribution efficiencies demonstrated in prior streaming studies~\cite{10.1145/3712676.3714450,usmani2025securingimmersive360video}.}

\subsection{Super-Resolution of Downsampled Datasets}
We implement SR as a supervised offset regression task using a Random Forest
model that maps sparse-point features to dense-point displacements. We chose Random Forest for its robustness to small training datasets, ability to model nonlinear feature relationships without extensive hyperparameter tuning, and interpretability, which is particularly valuable for analyzing feature contributions in geometric data. The model reconstructs dense point clouds from downsampled inputs by learning local geometric relationships. The overall pipeline comprises three stages: data preparation, model training, and inference. These stages establish the feature–offset mapping that enables the final SR step.

\subsubsection{Data Preparation} \label{subsubsec:data_preparation}
The SR task is formulated as a local geometric offset prediction problem, where the model learns to reconstruct dense geometry from a downsampled input. For each training pair, the downsampled (sparse) point cloud serves as the model input, while the corresponding full-resolution (dense) cloud provides supervision, i.e., each sparse point is associated with its nearest dense neighbors, from which the model learns the displacement vectors required to restore fine-grained structure. Each training sample thus encodes both absolute geometry and local density context:

\begin{itemize}
    \item For every sparse point $p_i = (x_i, y_i, z_i)$, its $K = 16$ nearest sparse neighbors are found using a k-dimensional tree (KD-tree). A KD-tree recursively partitions the 3D space along coordinate axes so that each node stores a splitting hyperplane and the points within its subspace. During a query, the tree prunes large portions of the search space by comparing distances to these hyperplanes, thus enabling near–logarithmic-time nearest-neighbor lookups~\cite{bentley1975multidimensional}. The mean neighbor distance $\bar{d}_K$ obtained from this search acts as a proxy for local sampling density: dense regions yield smaller $\bar{d}_K$, while sparse regions yield larger ones.
    
    \item The $M = 2$ closest dense points are then queried from the dense cloud using a separate KD-tree built on the dense coordinates. Their displacements from $p_i$ form offset targets $\Delta_i = (\Delta x, \Delta y, \Delta z)$, which describe the local geometric correction needed to upsample the sparse cloud. Selecting two nearest dense neighbors provides a fair balance between geometric accuracy and computational efficiency, i.e., the two offsets per sparse point capture local curvature while maintaining a manageable dataset size and a consistent $2\times$ upsampling ratio.
    
    \item Each feature vector is represented as $[x_i, y_i, z_i, \bar{d}_K, r]$, where $r \in \{1, 2\}$ denotes the \emph{rank} of the dense neighbor ($r{=}1$ for the nearest and $r{=}2$ for the second-nearest dense point). The rank term allows the Random Forest to learn how offset magnitudes vary with neighbor order. During inference, both ranks are evaluated for each sparse input. This yields two predicted dense offsets per point and thus a doubled output density. The corresponding label for each training sample is the offset $\Delta_i$.
\end{itemize}

This process is repeated for all available scenes and environments, producing paired feature-offset samples for each sparse-dense cloud pair. After constructing the dataset of all sparse-dense point correspondences across scenes, we merge them into a unified pool. Each sample retains its semantic tag, e.g., \textit{LivingRoom}, allowing us to later partition the data into domain-specific subsets for different training configurations. We discuss the dataset composition in Sect.~\ref{subsec:datset}.

\subsubsection{Model Training} \label{subsubsec:model_training}
After the subset extraction, two models are trained to analyze domain generalization. In the first configuration (\emph{Model~A}), the model is trained on one environment and evaluated on the other to assess cross-domain generalization. In the second configuration (\textit{Model~B}), data from both environments are combined to train a unified model, which is then evaluated on both domains to measure its ability to learn shared geometric representations and maintain robustness across environments. Each model consists of 300 decision trees with a maximum depth of~24.


\subsubsection{Inference}
During inference, the trained models are applied to unseen downsampled point clouds to reconstruct dense versions of each scene. For each sparse input file, the feature extraction process described in the data preparation stage is repeated: each sparse point $p_i = (x_i, y_i, z_i)$ is represented by its normalized coordinates, mean neighbor distance $\bar{d}_K$, and a rank value $r \in \{1,2\}$ corresponding to the two predicted dense offsets. Each feature vector is passed through the trained model, which predicts a displacement vector $\Delta_i = (\Delta x, \Delta y, \Delta z)$. The dense coordinates are then reconstructed as
\[
p_i' = p_i + \Delta_i.
\]
Because two ranks are evaluated per input point ($r=1,2$), each sparse point generates two corresponding dense predictions, resulting in a $2\times$ upsampled output cloud. \newtext{For higher factors (e.g., $4\times$ or $8\times$), the inference stage can be applied iteratively, using the previous dense output as the next sparse input.}

%% file: 4-Evaluation.tex
\section{Evaluation} \label{evaluation}


In this section, we evaluate the effects of downsampling on ABE computational performance and data size/bandwidth reduction, as well as the effectiveness of our SR model. 


\subsection{Point Cloud Dataset} \label{subsec:datset}
We utilize two point cloud datasets generated from RGB-D sensor captures, available in Open3D~\cite{Open3D}: \textit{LivingRoom} (56 files) and \textit{Office} (53 files). Each file is stored in the \texttt{.ply} format and contains spatial coordinates, surface normals, and colors. For simplicity, we remove normals and colors so that each point represents only its spatial position $(x, y, z)$. To ensure strict data separation between training and evaluation, 16 files are excluded from all data preparation and training.

As outlined in Sect.~\ref{subsubsec:data_preparation}, all sparse–dense correspondences generated from the downsampled and full-resolution point clouds are merged into a unified dataset. This dataset is stored using HDF5~(\texttt{.h5}), which provides chunked compression, fast random access, and scalable indexing for tens of millions of samples. Each record in the file contains the input feature vector, its corresponding offset label, the source domain tag (\textit{LivingRoom} or \textit{Office}), and the originating filename. 

After excluding the sixteen scenes reserved for evaluation, the remaining dense \texttt{.ply} files contain approximately 32.2~million points in total. Each file contains between 0.11--0.89~million points in the \textit{LivingRoom} domain and 0.11--0.63~million points in the \textit{Office} domain. From this unified dataset, we randomly sample 5~million pairs for training and 2~million pairs for testing in each configuration. These samples are used to train the Random Forest models described in Sect.~\ref{subsubsec:model_training}. The dense clouds serve as the ground-truth targets, while their downsampled counterparts provide the sparse inputs used to construct feature-offset pairs.

To evaluate the effectiveness of upsampling from downsampled point clouds and its impact on ABE computation, we select eight representative point cloud frames from the sixteen reserved scenes. These samples span a range of sizes and scene types, with four frames chosen from each dataset. Table~\ref{tab:heldout_samples} lists the specific point clouds used in our upsampling experiments.

\begin{table}[h]
\centering
\footnotesize
\caption{Frames selected for upsampling evaluation.}
\begin{tabular}{|p{1.5cm}|p{1.2cm}||p{1.5cm}|p{1.2cm}|}
\hline
\multicolumn{2}{|c||}{\textbf{LivingRoom}} & \multicolumn{2}{c|}{\textbf{Office}} \\
\hline
\textbf{Scene ID} & \textbf{Points} & \textbf{Scene ID} & \textbf{Points} \\
\hline
living-1 & 196,134 & office-14 & 282,072 \\
living-10 & 253,472 & office-22 & 333,504 \\
living-40 & 310,634 & office-37 & 380,898 \\
living-35 & 805,260 & office-28 & 430,814 \\
\hline
\end{tabular}
\label{tab:heldout_samples}
\end{table}

We first evaluate the predictive quality of the trained models on the evaluation dataset derived from these point clouds. Table~\ref{tab:offset_errors} summarizes the mean absolute error (MAE) and root mean square error (RMSE) of offset predictions for both configurations. The results show that both Random Forest models achieve sub-millimeter accuracy, indicating consistent learning performance across domains before geometric reconstruction is performed.

\begin{table}[t]
\centering
\footnotesize
\caption{Offset-level prediction errors of the Random Forest models on \newtext{evaluation dataset}.}
\label{tab:offset_errors}
\begin{tabular}{|p{0.9cm}|p{2.4cm}|p{0.9cm}|p{1.2cm}|p{1.2cm}|}
\hline
\textbf{Model} & \textbf{Training Mode} & \textbf{Test Set} & 
\textbf{MAE (mm)} & \textbf{RMSE (mm)} \\
\hline
\textbf{A} & Cross-domain (LivingRoom $\rightarrow$ Office) & Office & 0.327 & 0.771 \\
\textbf{B} & Mixed-domain (Living + Office) & Mixed & 0.322 & 0.768 \\
\hline
\end{tabular}
\end{table}

\begin{figure}[htbp]
  \centering
  \includegraphics[width=0.24\linewidth]{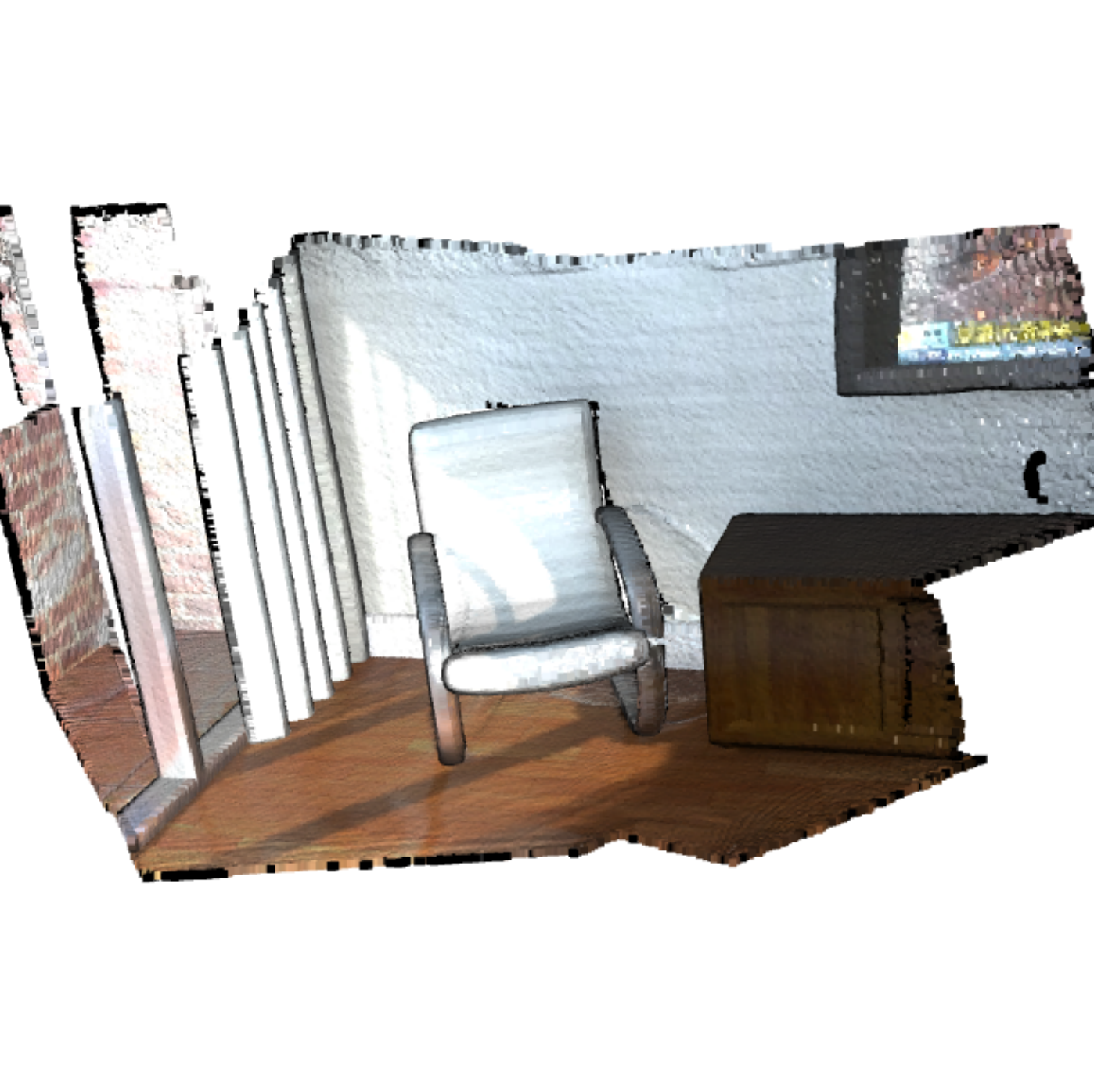}
  \includegraphics[width=0.24\linewidth]{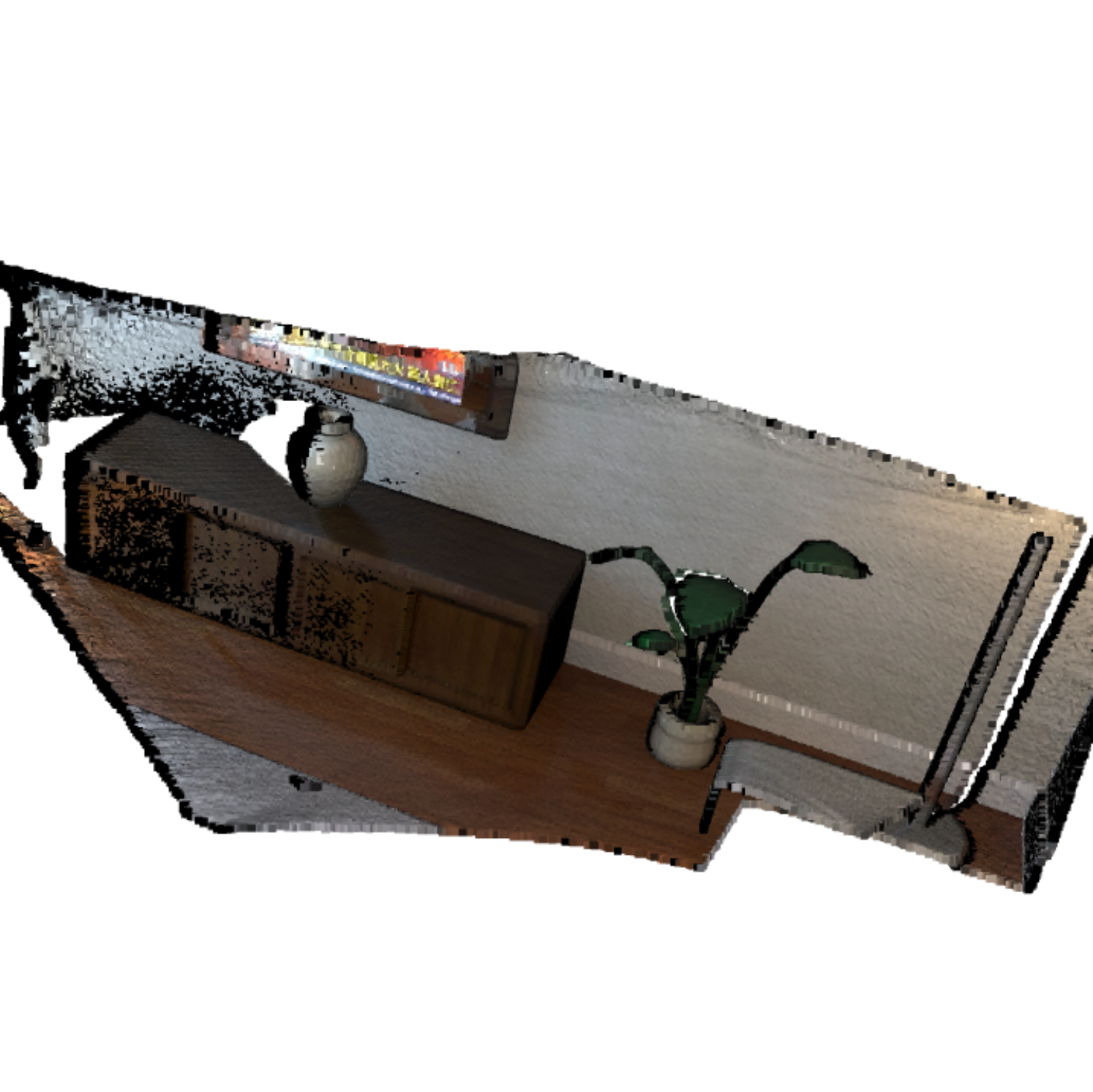}
  \includegraphics[width=0.24\linewidth]{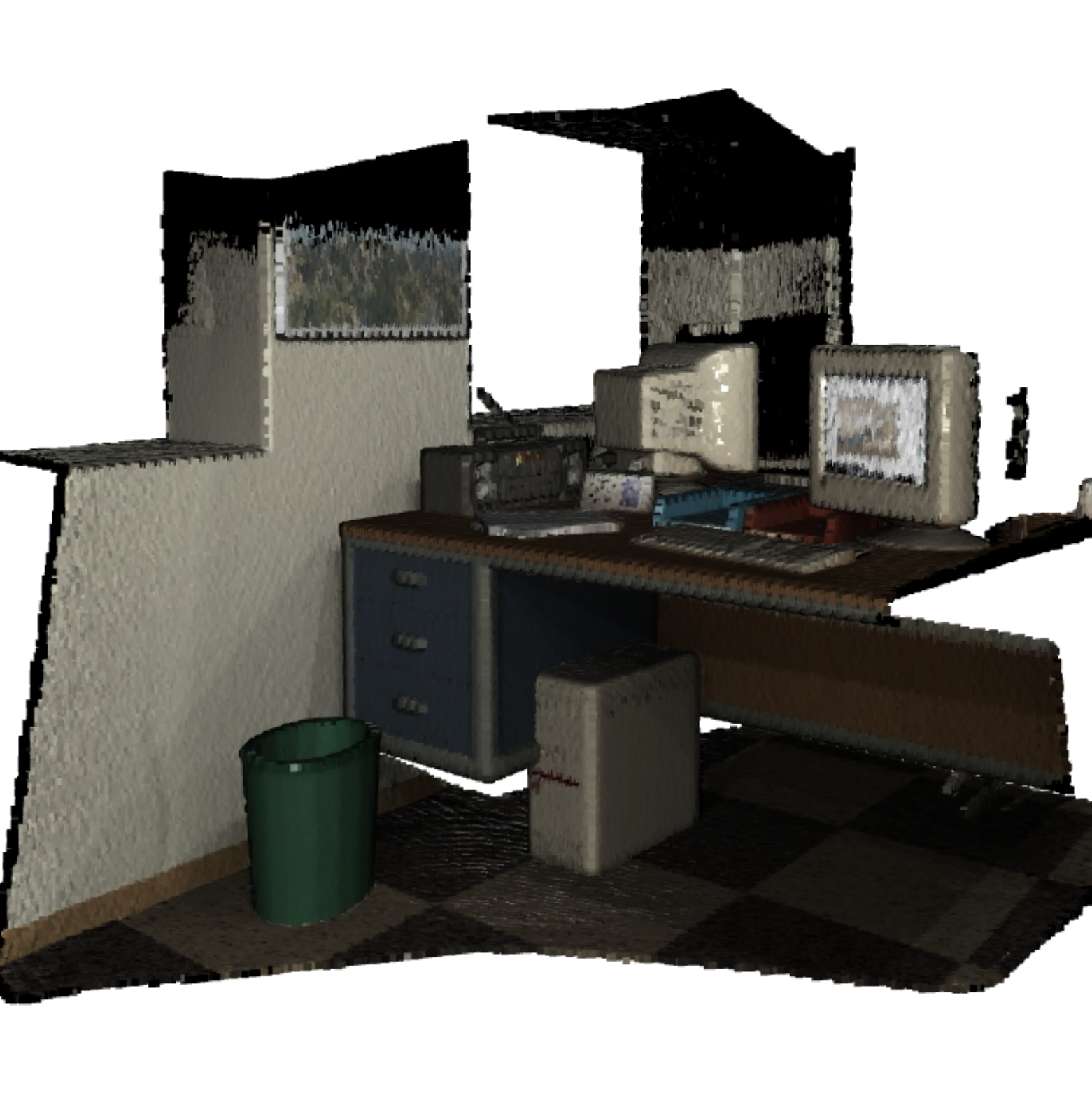}
  \includegraphics[width=0.24\linewidth]{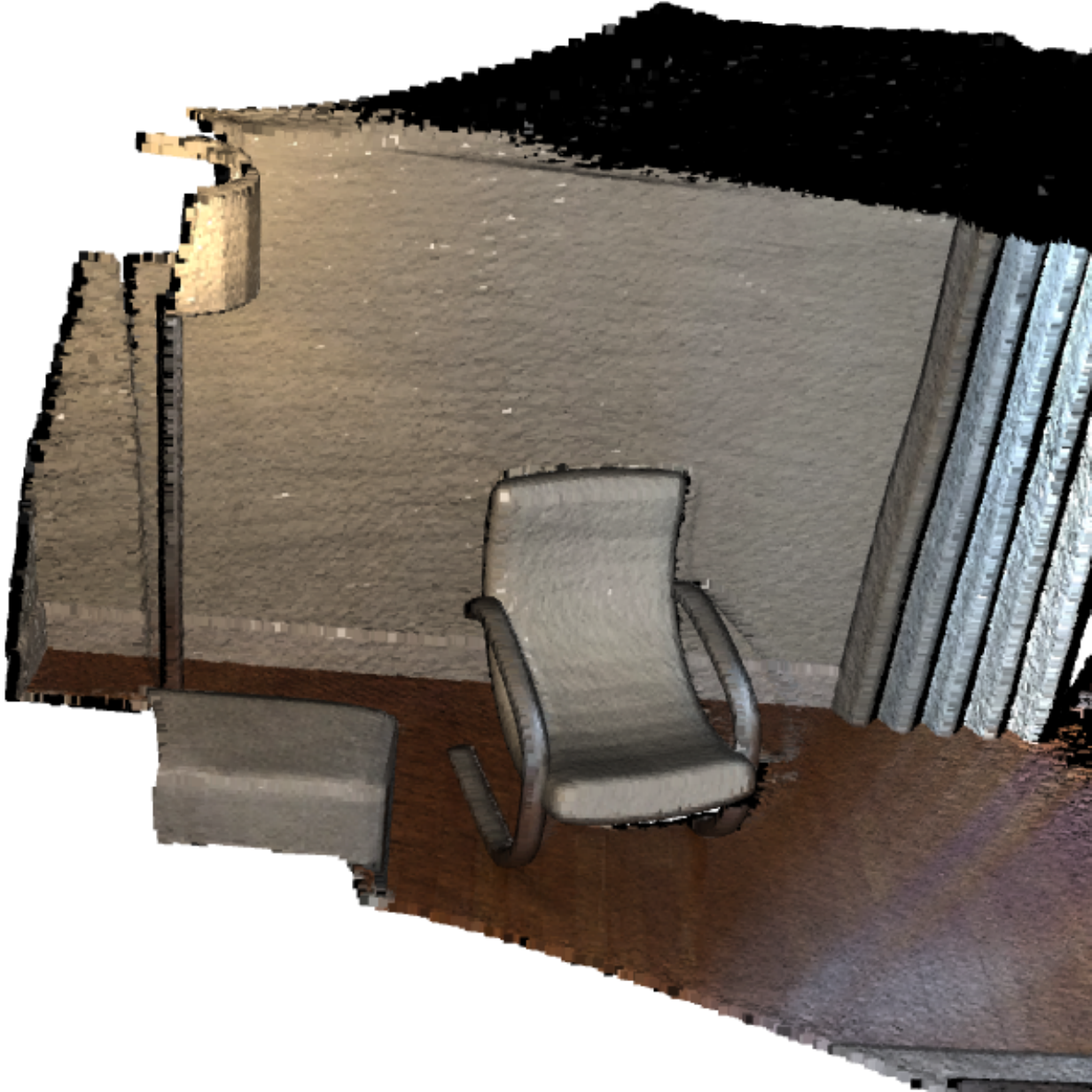}
    \includegraphics[width=0.24\linewidth]{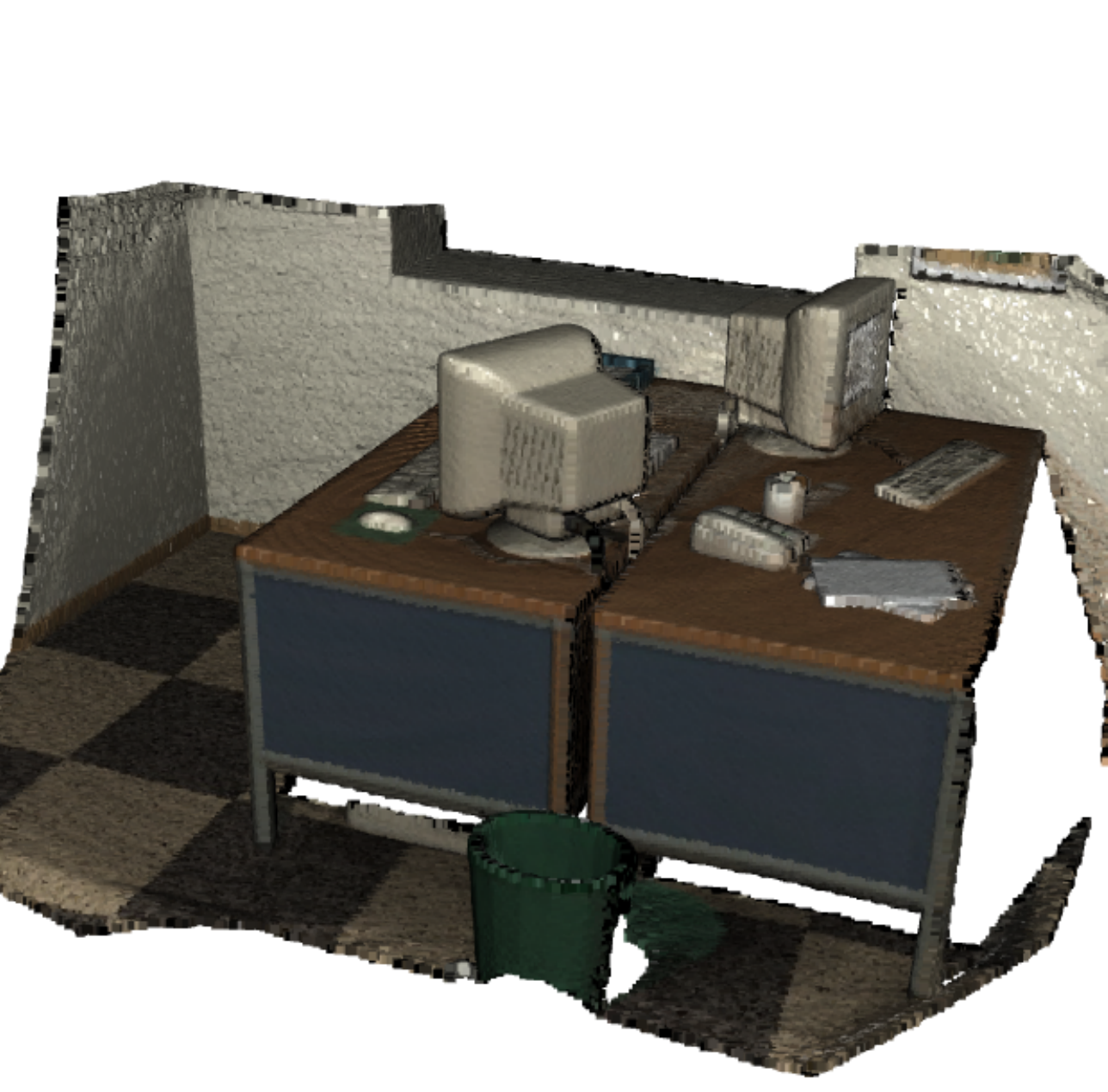}
  \includegraphics[width=0.24\linewidth]{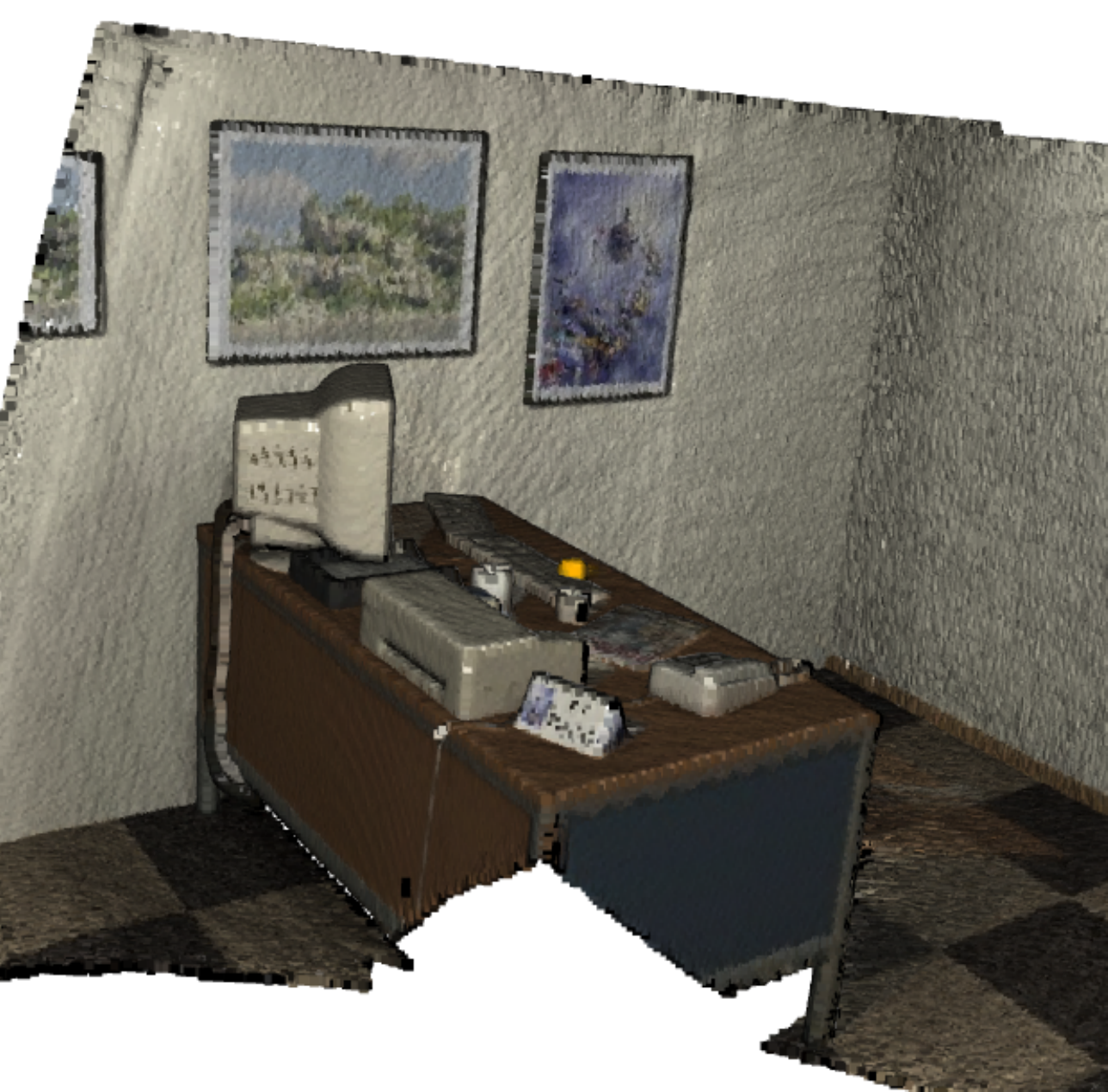}
  \includegraphics[width=0.24\linewidth]{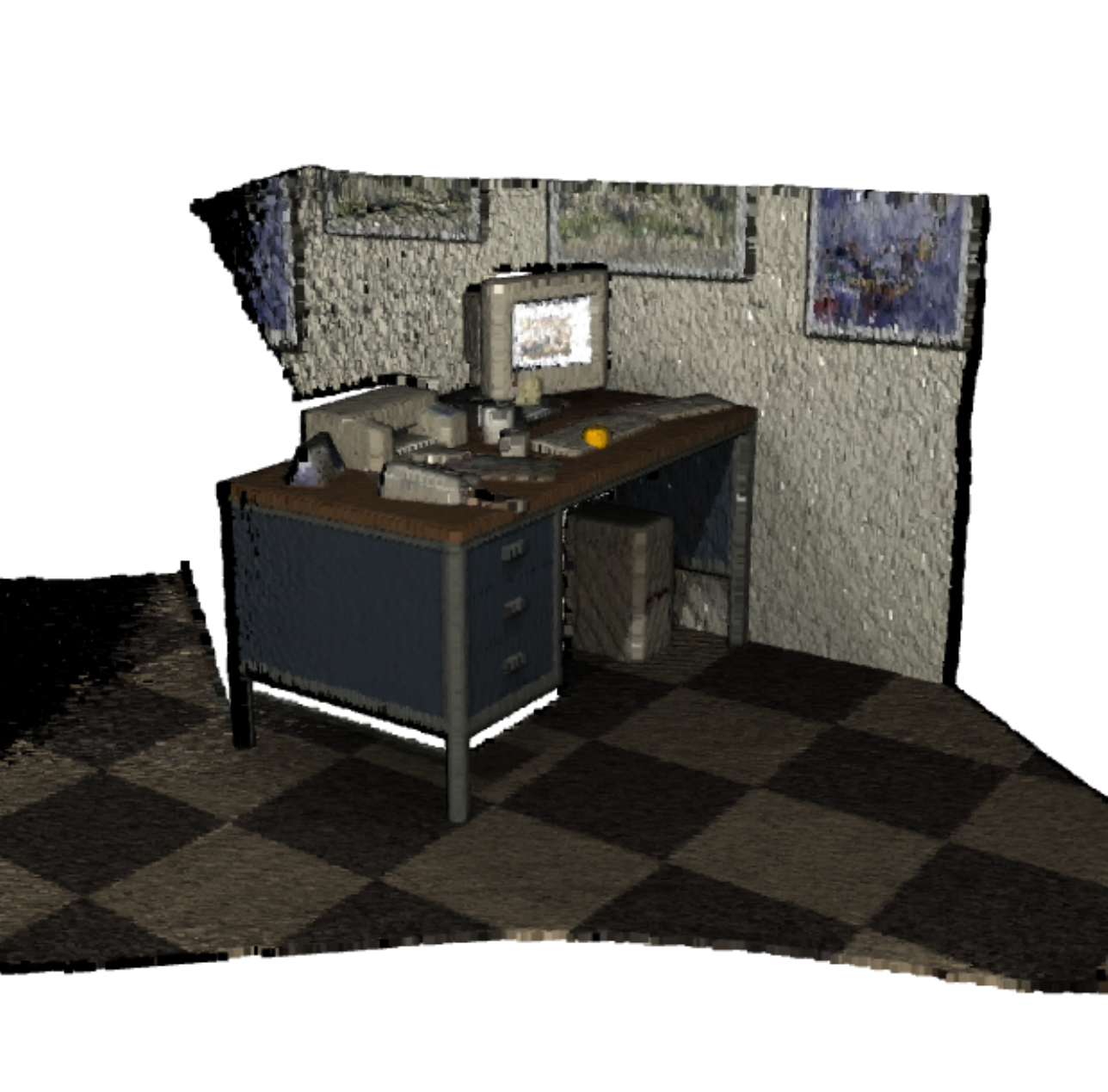}
  \includegraphics[width=0.24\linewidth]{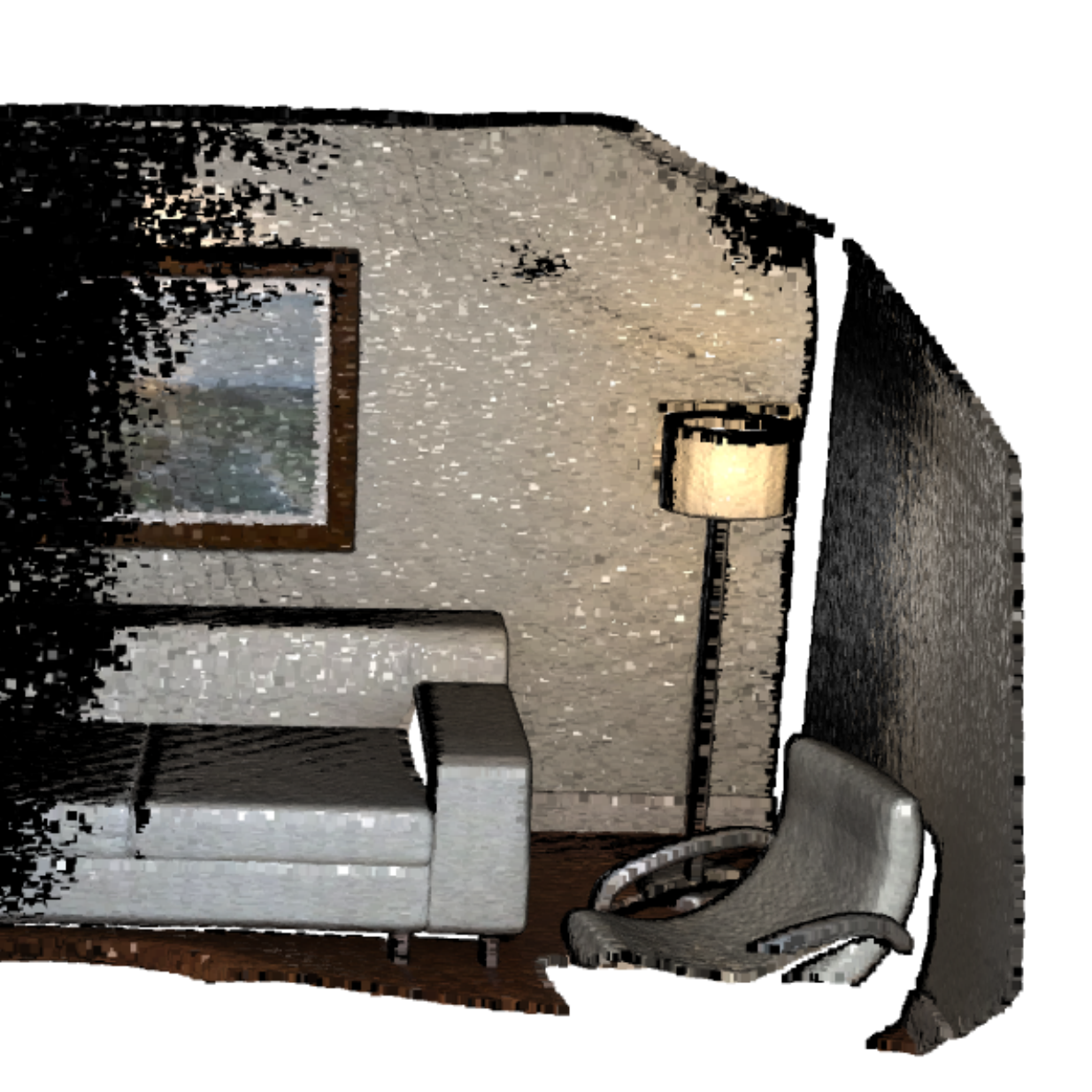}
  \caption{Visuals of the original point clouds: 196k, 253k, 282k, 310k, 333k, 380k, 430k, 805k.}
  \label{fig:pointcloud_examples}
\end{figure}

\newtext{The} selection captures a diverse range of point cloud sizes and scene types, allowing for a meaningful comparative analysis. Visual examples of these point clouds are presented in Fig.~\ref{fig:pointcloud_examples}. These are the original visuals before removing the surface normals and colors.

\subsection{Impact of Downsampling on ABE and Data Size}\label{abe-size-eval}
We evaluate the impact of downsampling on both ABE computational performance and data size for point cloud frames. Specifically, we measure the encryption and decryption times, along with the corresponding reductions in file size and bandwidth required for real-time streaming.


\newtext{In our experiments, all point clouds are encrypted at the full $X, Y, Z$ coordinate granularity using ABE. Encryption is performed on the content provider’s side and involves writing encrypted frames to disk, as the pre-encrypted versions are stored on the origin server for streaming. Decryption, in contrast, occurs on the client side, typically on resource-constrained devices that may also perform SR to restore the original spatial resolution, adding computational overhead. To minimize this overhead, decryption is performed in-memory without disk writes, allowing frames to be decrypted, upsampled, and rendered on-the-fly without persistent storage.}

This evaluation uses eight representative point clouds (Sect.~\ref{subsec:datset}) and compares downsampled versions containing 50\%, 25\%, and 12.5\% of the original points against the baseline 100\% (full-resolution) point clouds. This setup allows us to quantify both the computational savings in ABE operations and the reduction in data size and bandwidth demand achieved through downsampling.

\begin{figure*}
    \centering
    \includegraphics[width=\linewidth]{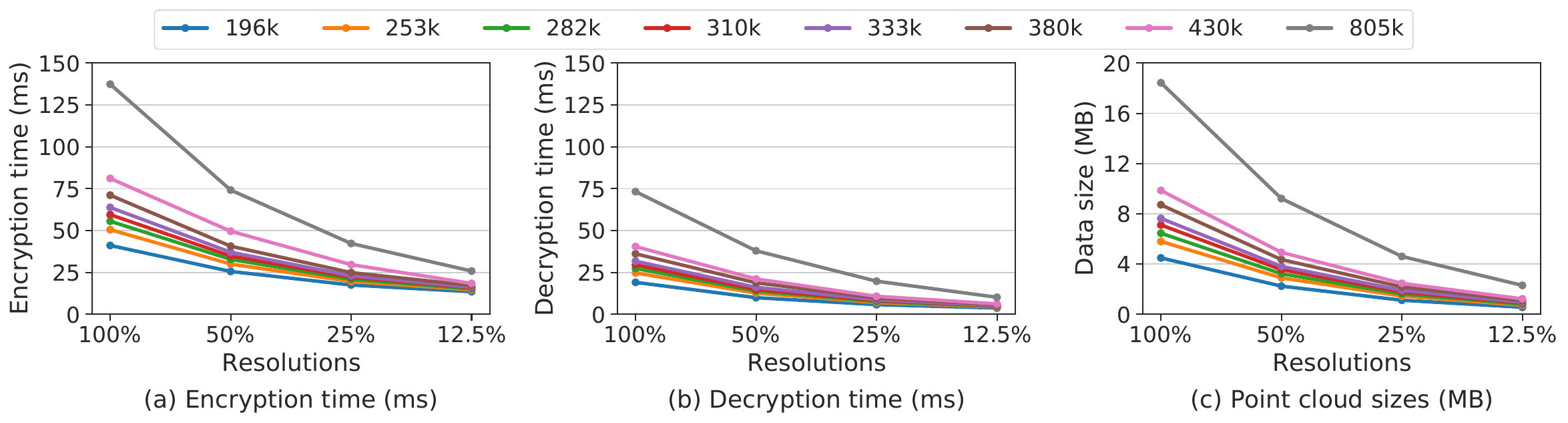}
    \caption{Encryption and decryption times, and encrypted data sizes across various resolutions for different point cloud sizes.}
    \label{fig:pcs-enc-dec-size-eval}
\end{figure*}

\paragraph{Encryption and Decryption Time Analysis}
Figures~\ref{fig:pcs-enc-dec-size-eval}(a) \& (b) show the encryption and decryption times (in milliseconds) for each point cloud at different resolutions. The results indicate that, across all point clouds, downsampling from 100\% to 50\% reduces encryption time by an average of 41\%. At 25\% of the points, the reduction increases to approximately 63\%, and for 12.5\%, it reaches nearly 75\% on average.

A similar trend is observed for decryption times: about 49\% reduction at 50\% resolution, an average of 72\% at 25\%, and up to 86\% at 12.5\%. These reductions closely follow the decrease in the number of points (e.g., 25\% resolution corresponds to 75\% fewer points), suggesting that \textit{decryption time scales almost linearly with the number of points}. The savings are more pronounced during decryption than encryption, which is a key advantage


\paragraph{Data Size Analysis}
Fig.~\ref{fig:pcs-enc-dec-size-eval}(c) illustrates the data sizes (in megabytes) for each point cloud at different resolutions. These sizes include the overhead introduced by ABE. The results demonstrate a \textit{linear decrease in size with lower resolutions}—for instance, a 49.99\% reduction at 50\% resolution, 74.99\% at 25\%, and continuing proportionally for lower levels. This linear reduction in data size also directly translates to a proportional decrease in bandwidth usage during transmission.

In summary, downsampling provides complementary benefits within streaming systems, significantly reducing bandwidth requirements during transmission and decreasing client-side decryption latency by minimizing the amount of data processed. Furthermore, the reduction in encryption time on the content provider’s side offers additional advantages, particularly in live streaming scenarios where point cloud frames must be encrypted and delivered in real time.

\subsection{Evaluation of Super-Resolution Model}
We evaluate the performance of our AI/ML-based upsampling models by reconstructing high-resolution point clouds from their downsampled counterparts (Sect.~\ref{subsec:datset}). To quantitatively assess reconstruction fidelity, we employ two standard spatial similarity metrics: the Chamfer Distance (CD)~\cite{9879253} and the Hausdorff Distance (HD)~\cite{10.1145/3503161.3548384}. The CD measures the average bidirectional nearest-neighbor distance between the reconstructed and reference point clouds, while the HD captures the maximum of these minimum distances, reflecting the worst-case geometric deviation. Lower values of both metrics indicate higher reconstruction fidelity and closer geometric alignment with the original dense point cloud.

In our experiments, three downsampled resolutions—12.5\%, 25\%, and 50\% of the original point density—are used as input. Each is upsampled back to full 100\% resolution using the trained Random Forest models~(Sect.~\ref{subsubsec:model_training}). Lower-resolution inputs (12.5\% and 25\%) are incrementally upsampled in two or more stages (e.g., 12.5\%~$\rightarrow$~25\%~$\rightarrow$~50\%~$\rightarrow$~100\%). Regardless of the intermediate stages, evaluation is performed by comparing the final reconstructed 100\% cloud with its corresponding original 100\% dense version. We also record the inference time required for each reconstruction.

\begin{figure*}[t]
    \centering
    \begin{subfigure}[b]{0.33\textwidth}
        \centering
        \includegraphics[width=\textwidth]{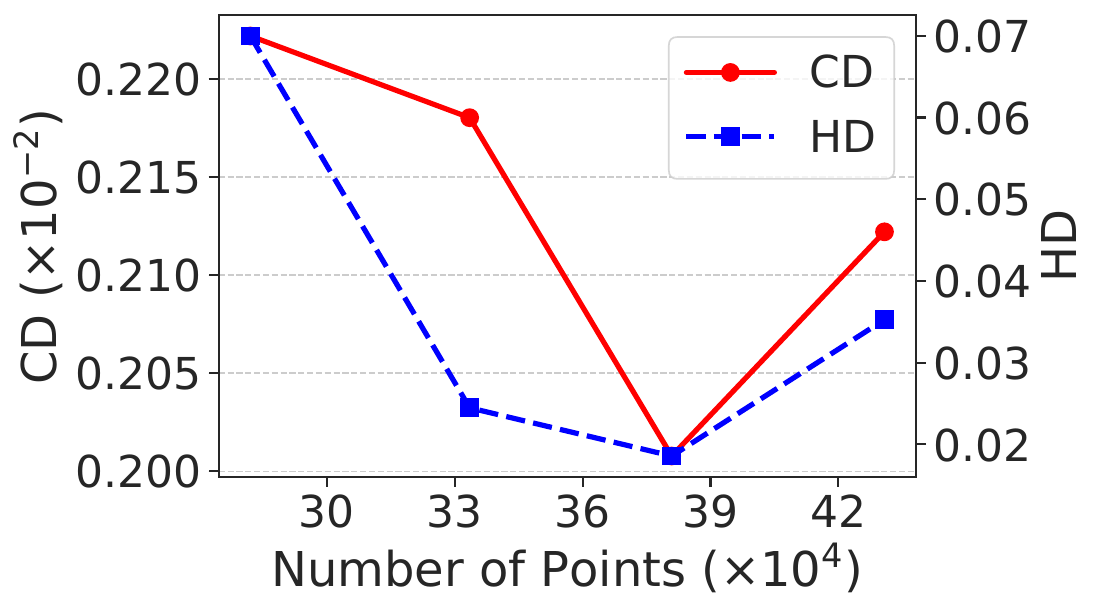}
        \caption{50\%~$\rightarrow$~100\% upsampling (2$\times$).}
        \label{fig:cd_hd_modelA_scale2x}
    \end{subfigure}\hfill
    \begin{subfigure}[b]{0.33\textwidth}
        \centering
        \includegraphics[width=\textwidth]{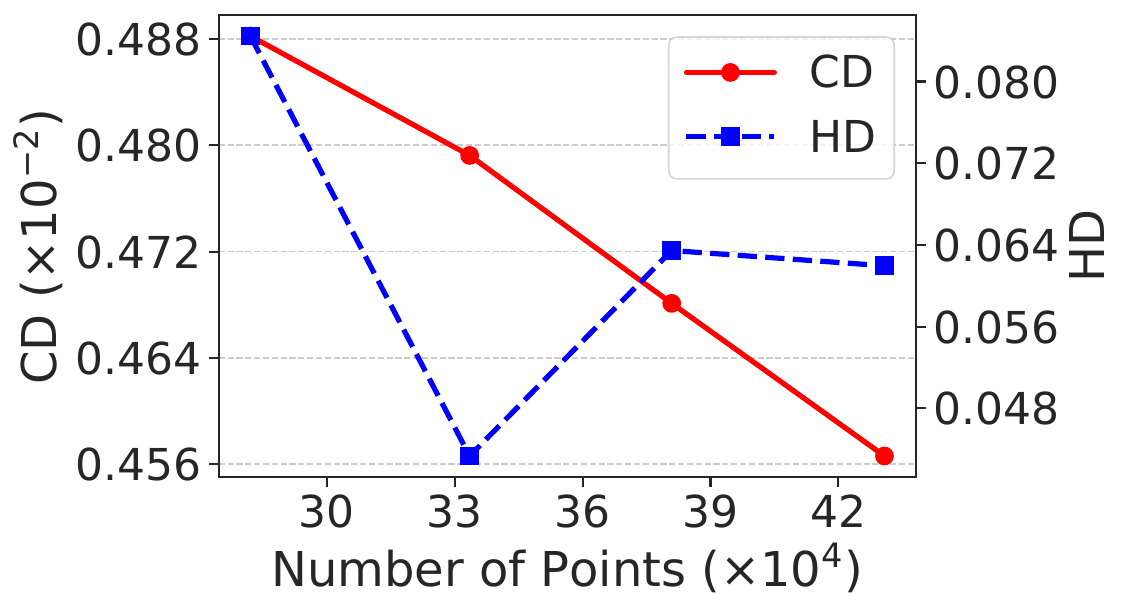}
        \caption{25\%~$\rightarrow$~100\% upsampling (4$\times$).}
        \label{fig:cd_hd_modelA_scale4x}
    \end{subfigure}\hfill
    \begin{subfigure}[b]{0.33\textwidth}
        \centering
        \includegraphics[width=\textwidth]{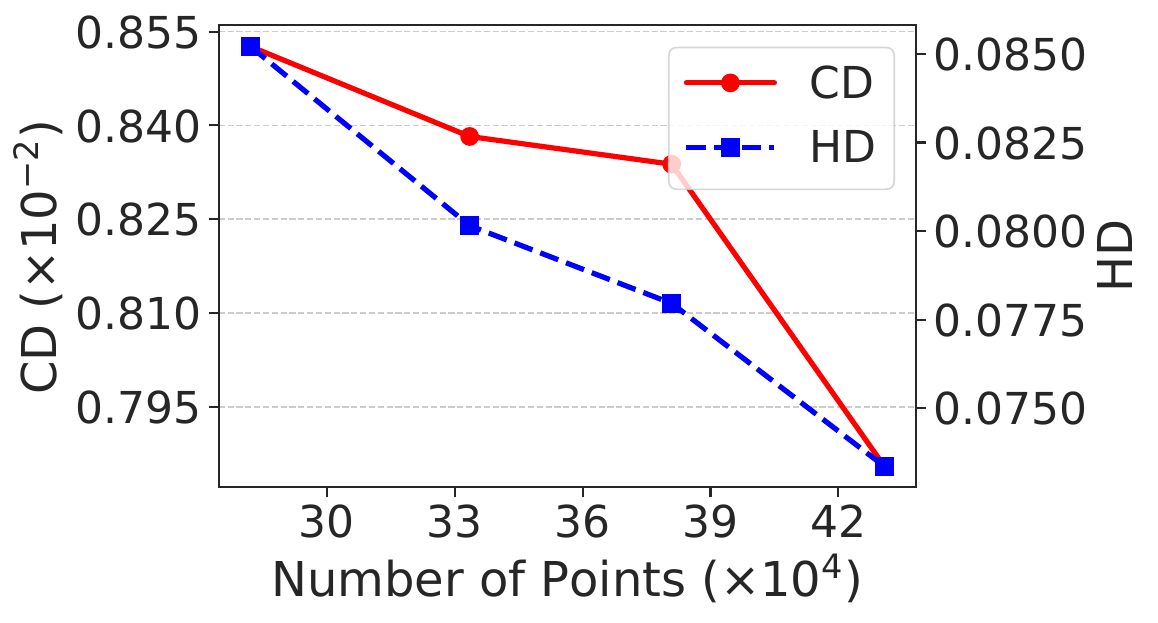}
        \caption{12.5\%~$\rightarrow$~100\% upsampling (8$\times$).}
        \label{fig:cd_hd_modelA_scale8x}
    \end{subfigure}

    \caption{Chamfer and Hausdorff distances for Model~A (cross-domain) under different input densities upsampled to 100\%.}
    \label{fig:cd_hd_modelA_all}
\end{figure*}

\begin{figure*}[t]
    \centering
    \begin{subfigure}[b]{0.33\textwidth}
        \centering
        \includegraphics[width=\textwidth]{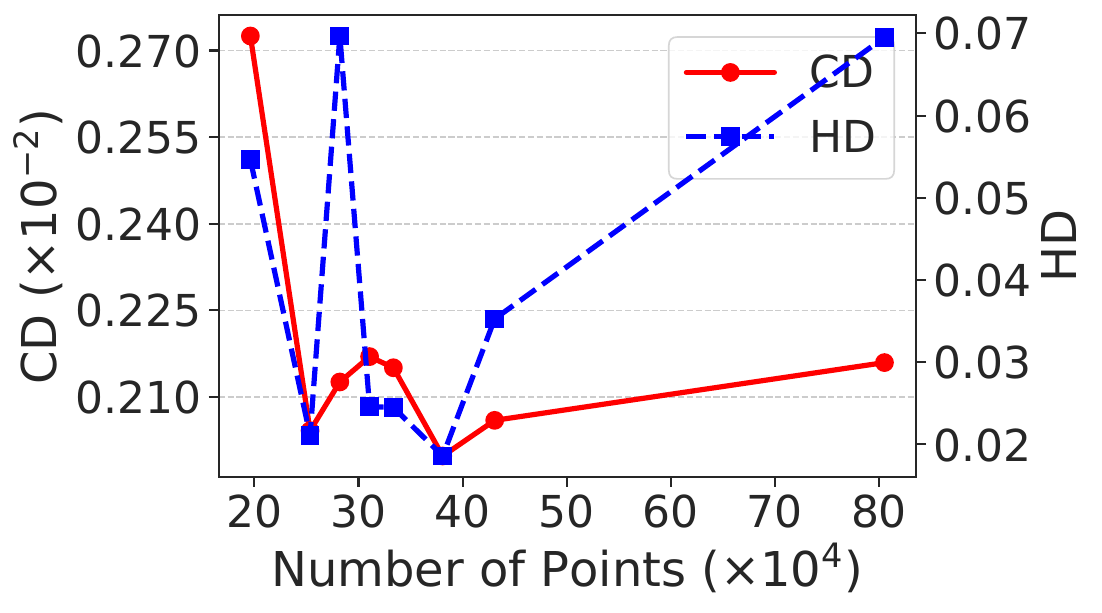}
        \caption{50\%~$\rightarrow$~100\% upsampling (2$\times$).}
        \label{fig:cd_hd_modelC_scale2x}
    \end{subfigure}\hfill
    \begin{subfigure}[b]{0.33\textwidth}
        \centering
        \includegraphics[width=\textwidth]{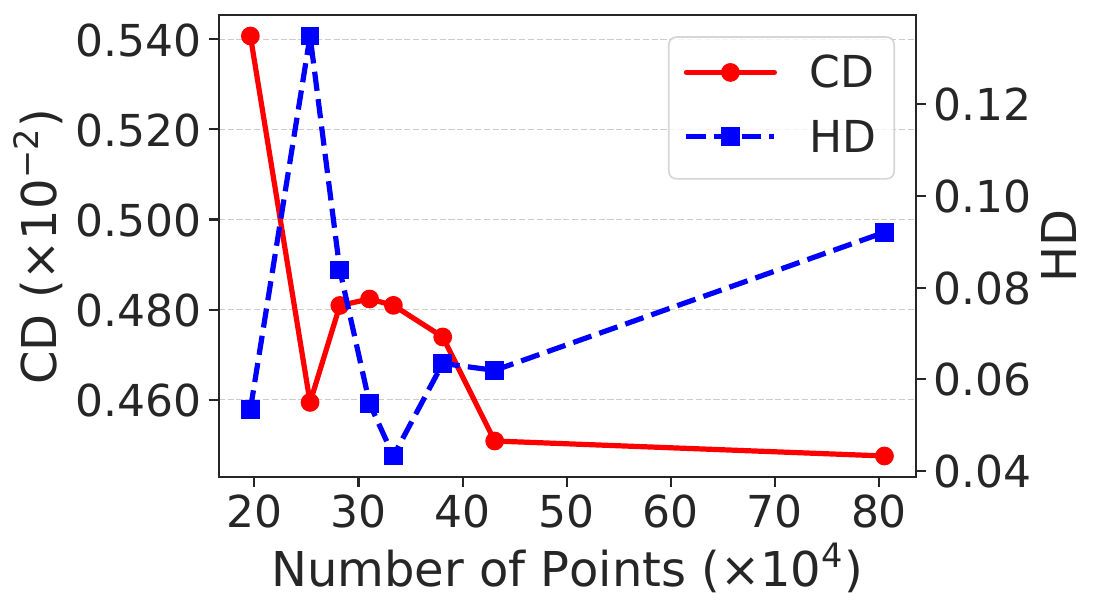}
        \caption{25\%~$\rightarrow$~100\% upsampling (4$\times$).}
        \label{fig:cd_hd_modelC_scale4x}
    \end{subfigure}\hfill
    \begin{subfigure}[b]{0.33\textwidth}
        \centering
        \includegraphics[width=\textwidth]{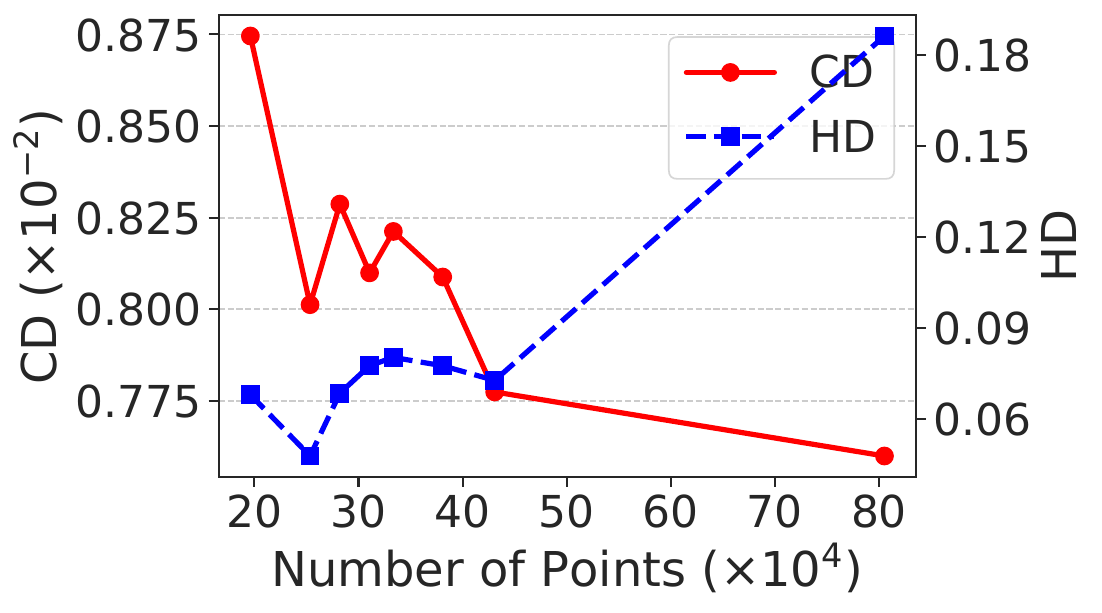}
        \caption{12.5\%~$\rightarrow$~100\% upsampling (8$\times$).}
        \label{fig:cd_hd_modelB_scale8x}
    \end{subfigure}

    \caption{Chamfer and Hausdorff distances for Model~B (mixed-domain) under different input densities upsampled to 100\%.}
    \label{fig:cd_hd_modelC_all}
\end{figure*}

\begin{figure}
    \centering
    \includegraphics[width=\linewidth]{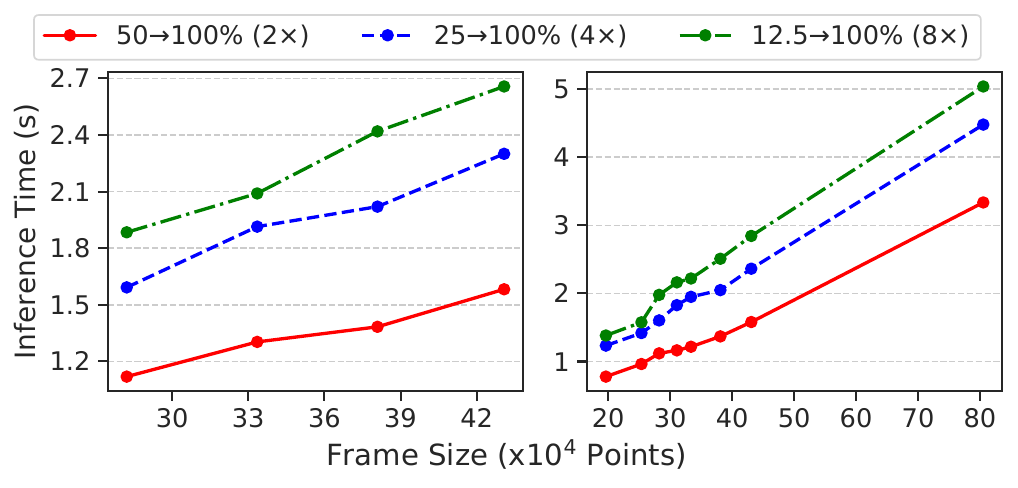}
    \caption{Inference time per frame for Models~A (\emph{left}, cross-domain) and~B (\emph{right}, mixed-domain) 
    under 12.5\%, 25\%, and 50\% input densities upsampled to 100\%.}
    \label{fig:inference_models_A_C}
\end{figure}

Figures~\ref{fig:cd_hd_modelA_all} and~\ref{fig:cd_hd_modelC_all} show the CD and HD for the two Random Forest configurations across all three input resolutions. The 25\%~$\rightarrow$~100\% and 12.5\%~$\rightarrow$~100\% upsampling runs show slightly higher error values than the 50\%~$\rightarrow$~100\% runs because they start from fewer input samples. With fewer points, the model has less geometric context to rely on, leading to larger local interpolation gaps and slightly higher distances. However, both configurations maintain consistent Chamfer and Hausdorff ranges across the three scales, and demonstrate stable reconstruction quality under varying degrees of sparsity.

When comparing the two models, the mixed-domain configuration (\emph{Model~B}) follows a pattern similar to the cross-domain configuration (\emph{Model~A}), with comparable error ranges across the same set of held-out files. Within this setup, combining \textit{LivingRoom} and \textit{Office} data during training did not adversely affect reconstruction accuracy, and both models maintained consistent Chamfer and Hausdorff values across scenes, and demonstrate stable generalization across environments.

Fig.~\ref{fig:inference_models_A_C} summarizes the per-frame inference times of Models~A and~B across different input resolutions. The inference runtime of the Random-Forest-based SR model remains lightweight across frames. Direct upsampling from 50\% to 100\% density completes in under two seconds for most frames, with smaller scenes finishing within one second and the largest (with around 800K points) taking about three seconds. Multi-stage 2× chaining introduces additional latency because each intermediate stage reconstructs neighborhood structures and re-extracts features for an increasingly dense point set, compounding the cost of KD-tree queries and per-point predictions. An end-to-end model trained to map directly from coarse inputs (e.g., 25\% or 12.5\%) to full resolution could avoid such redundant feature computations and reduce overall runtime. For mixed-reality pipelines, frame latencies below roughly 20-50~ms are typically considered real-time. Bridging this gap, particularly on resource-constrained clients, would require architectural optimizations such as reusing neighborhood structures across consecutive frames and offloading feature computation to nearby edge servers~\cite{alshahrani2020efficient,zhang2022end}. We plan to investigate system-level optimizations for our SR framework in future work.

\subsection{Evaluation of Streaming Latency with Downsampling}\label{subsec:streaming-latency}


To evaluate streaming latency, we conducted a simulation on CloudLab~\cite{duplyakin2019cloudlab} using two physical nodes: a server hosting pre-encrypted point cloud frames via Apache HTTP
and a client running Apache Benchmark (ab)
to generate load. The experiment measured the effect of frame resolution and decryption overhead on volumetric streaming latency. We aimed to emulate a realistic streaming setup where 10 clients watch a 30 FPS volumetric video for 600 seconds, resulting in a total of 180,000 HTTP requests (10 x 30 × 600). This workload was issued with concurrency level set to 10, effectively simulating 10 clients fetching frames in parallel.

As the volumetric content, we used a point cloud the dataset (Table~\ref{tab:heldout_samples} \textit{office-28} with \textit{430K} points). To model a full-length video, the same frame was repeatedly requested imitating streaming of a point cloud sequence. We conducted four experiments using different resolutions of the point cloud: 100\% 
, 50\%, 25\%, and 12.5\%, obtained by downsampling the original point cloud. Each of these versions was pre-encrypted using ABE, such that the network payload included encryption metadata, ensuring a realistic evaluation of actual streaming conditions with security overhead.

\begin{figure}
    \centering
    \includegraphics[width=\linewidth]{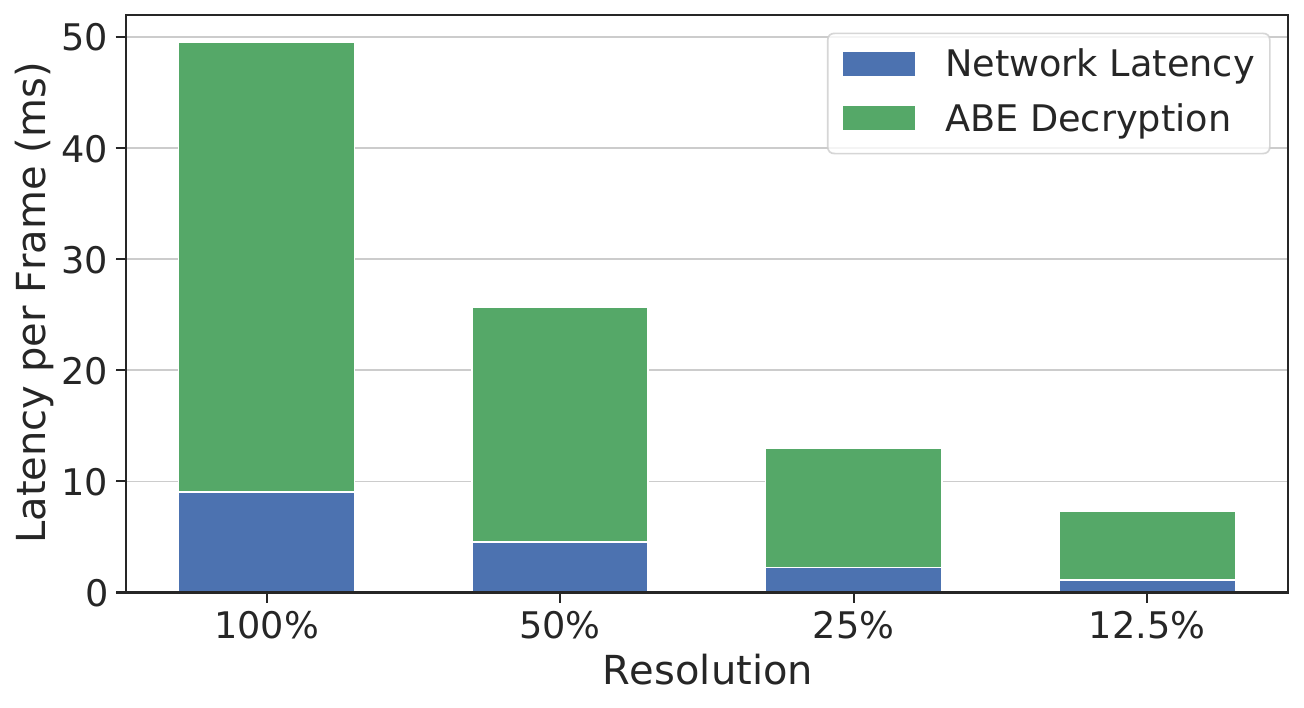}
    \caption{Streaming latency under different resolutions. Blue segment shows network latency, and green segment shows additional ABE decryption time.}
    \label{fig:streaming-latency}
\end{figure}




\newtext{We measured total latency by combining network transmission time and client-side ABE decryption delay. Network latency, obtained using the mean time per HTTP request from Apache Benchmark, represents the average time to serve a single point cloud frame across all clients. As shown in Fig.\ref{fig:streaming-latency}, network latency (blue bars) is highest for the full-resolution (100\%) frames at around 9~ms and decreases nearly linearly with downsampling—about 4.5~ms at 50\%, with 25\% and 12.5\% following the same trend. When ABE decryption time is included, the overall latency (green bars) exhibits the same linear behavior, confirming that downsampling combined with ABE provides an efficient and scalable means to reduce end-to-end delay in volumetric point cloud streaming.}

%% file: 5-Discussion.tex
\section{Discussion and Future Work}

Downsampling reduces bandwidth requirements during transmission and lowers computational overhead during decryption. These benefits increase as the sampling resolution decreases. However, this comes at the cost of reduced geometric similarity, as reflected by higher CD and  HD values, and longer inference times during upsampling. For instance, reconstructing the original 100\% resolution from a 12.5\% downsampled version introduces greater dissimilarity and higher inference latency, which may offset or diminish the gains achieved through downsampling. This optimal point varies depending on factors such as network conditions, device capabilities, and model complexity. We will explore this in a future work.


\newtext{Although streaming-side evaluations are beyond the scope of this paper, they remain an important direction for future work. In particular, we plan to extend this work toward an end-to-end implementation and evaluation under real-world network conditions, enabling measurement of throughput, latency, and quality trade-offs across the full transmission, decryption, and reconstruction pipeline.}

\newtext{We plan to explore joint optimization of computation and data transfer through adaptive downsampling at the source. We aim to formulate this optimization using classical operations research or convex optimization techniques, or learn it through data-driven approaches such as machine learning or reinforcement learning.}

\newtext{Finally, we plan to investigate architectural optimizations for resource-constrained clients, such as reusing spatial relationships across consecutive frames and offloading computationally intensive inference steps to nearby edge nodes, for optimizing inference efficiency.}

%% file: 6-Conclusion.tex
\section{Conclusion}

\newtext{In this work, we presented a novel framework that integrates point cloud downsampling and ML-based super-resolution with
ABE, introducing a security-aware perspective to efficient volumetric data processing. Our results demonstrate that downsampling not only reduces point cloud data size—and consequently bandwidth \& latency requirements but also lowers the computational overhead of ABE encryption and decryption. 
Our proposed AI/ML model effectively reconstructs full-resolution point clouds with minimal error and modest inference time. The observed linear relationship between downsampling level and resource savings 
shows that the approach can scale well and is practical for immersive media applications.
Although we focus on computational performance and reconstruction accuracy, we will extend this framework to real-time streaming scenarios to further validate its effectiveness under dynamic network conditions.}

